\documentclass[reprint,
superscriptaddress,
 amsmath,amssymb,
 aps,
prb,
floatfix,
]{revtex4-2}
\usepackage{hyperref}
\usepackage{graphicx}
\usepackage{dcolumn}
\usepackage{bm}
\usepackage{cleveref}
\usepackage{color}
\usepackage{comment}
\usepackage{float}
\newcommand{\ve}[1]{\boldsymbol{#1}}
\usepackage{booktabs}
\usepackage{multirow}
\usepackage{makecell}
\usepackage{color, colortbl}
\usepackage[dvipsnames]{xcolor}

\definecolor{Gray}{gray}{0.95}
\newcolumntype{g}{>{\columncolor{Gray}}c}

\preprint{APS/123-QED}
\begin{document}
\title{Imaging the transition from diffusive to Landauer resistivity dipoles}

\author{Serhii Kovalchuk}
\affiliation{Peter Gr\"unberg Institut (PGI-3), Forschungszentrum J\"ulich, 52425 J\"ulich, Germany}
\affiliation{J\"ulich Aachen Research Alliance (JARA), Fundamental of Future Information Technology, 52425 J\"ulich, Germany}
\affiliation{Institute of Experimental Physics, University of Wrocław, pl. M. Borna 9, 50-204, Wrocław, Poland}

\author{David K\"ampfer}
\affiliation{Peter Gr\"unberg Institut (PGI-3), Forschungszentrum J\"ulich, 52425 J\"ulich, Germany}
\affiliation{J\"ulich Aachen Research Alliance (JARA), Fundamental of Future Information Technology, 52425 J\"ulich, Germany}
\affiliation{Lehrstuhl f\"ur Experimentalphysik IV A, RWTH Aachen University, Otto-Blumenthal-Straße, 52074 Aachen,Germany}

\author{Jonathan K. Hofmann}
\affiliation{Peter Gr\"unberg Institut (PGI-3), Forschungszentrum J\"ulich, 52425 J\"ulich, Germany}
\affiliation{J\"ulich Aachen Research Alliance (JARA), Fundamental of Future Information Technology, 52425 J\"ulich, Germany}
\affiliation{Lehrstuhl f\"ur Experimentalphysik IV A, RWTH Aachen University, Otto-Blumenthal-Straße, 52074 Aachen,Germany}

\author{Timofey Balashov}
\affiliation{Peter Gr\"unberg Institut (PGI-3), Forschungszentrum J\"ulich, 52425 J\"ulich, Germany}
\affiliation{J\"ulich Aachen Research Alliance (JARA), Fundamental of Future Information Technology, 52425 J\"ulich, Germany}
\affiliation{Lehrstuhl f\"ur Experimentalphysik II A, RWTH Aachen University, Otto-Blumenthal-Straße, 52074 Aachen,Germany}

\author{Vasily Cherepanov}
\affiliation{Peter Gr\"unberg Institut (PGI-3), Forschungszentrum J\"ulich, 52425 J\"ulich, Germany}
\affiliation{J\"ulich Aachen Research Alliance (JARA), Fundamental of Future Information Technology, 52425 J\"ulich, Germany}

\author{Bert Voigtländer}
\affiliation{Peter Gr\"unberg Institut (PGI-3), Forschungszentrum J\"ulich, 52425 J\"ulich, Germany}
\affiliation{J\"ulich Aachen Research Alliance (JARA), Fundamental of Future Information Technology, 52425 J\"ulich, Germany}
\affiliation{Lehrstuhl f\"ur Experimentalphysik IV A, RWTH Aachen University, Otto-Blumenthal-Straße, 52074 Aachen,Germany}

\author{Ireneusz Morawski}
\affiliation{Institute of Experimental Physics, University of Wrocław, pl. M. Borna 9, 50-204, Wrocław, Poland}

\author{F. Stefan Tautz}
\affiliation{Peter Gr\"unberg Institut (PGI-3), Forschungszentrum J\"ulich, 52425 J\"ulich, Germany}
\affiliation{J\"ulich Aachen Research Alliance (JARA), Fundamental of Future Information Technology, 52425 J\"ulich, Germany}
\affiliation{Lehrstuhl f\"ur Experimentalphysik IV A, RWTH Aachen University, Otto-Blumenthal-Straße, 52074 Aachen,Germany}

\author{Felix L\"upke}
\email{f.luepke@fz-juelich.de}
\affiliation{Peter Gr\"unberg Institut (PGI-3), Forschungszentrum J\"ulich, 52425 J\"ulich, Germany}
\affiliation{J\"ulich Aachen Research Alliance (JARA), Fundamental of Future Information Technology, 52425 J\"ulich, Germany}
\affiliation{II. Physikalisches Institut, Universität zu K\"oln, Z\"ulpicher Straße 77, 50937 K\"oln, Germany}

\date{\today}

\begin{abstract}
A point-like defect in a uniform current-carrying conductor induces a dipole in the electrochemical potential, which counteracts the original transport field. 
If the mean free path of the carriers is much smaller than the size of the defect, the dipole results from the purely diffusive motion of the carriers around the defect. 
In the opposite limit, ballistic carriers scatter from the defect---for this situation, Rolf Landauer postulated the emergence of residual resistivity dipoles that are independent of the defect size and thus impose a fundamental limit on the resistance of the parent conductor. 
Here, we study resistivity dipoles around holes of different sizes in two-dimensional Bi films on Si(111). 
Using scanning tunneling potentiometry to image the dipoles, we find a crossover from linear to constant scaling behavior of their amplitudes with defect size, manifesting the transition from diffusive to Landauer dipoles.
The extracted parameters of the transition allow us to estimate the Fermi wave vector and the carrier mean free path in our Bi films.
\end{abstract}

\maketitle

\section{Introduction}
When a directed current flows through a conductor, the scattering of charge carriers at defects leads to charge accumulation in front of the defect and charge depletion behind the defect (with respect to the overall current direction). This results in a local electric dipole that counteracts the overall transport field \cite{Landauer1957, Feenstra1998}, causing a macroscopically observable increase in electrical resistance \cite{Feenstra1998, Luepke2017}. For point-like (quasi zero-dimensional) defects, scattering follows either diffusive or ballistic transport models, depending on whether the size of the defect is larger or smaller than the mean free path of the carriers. While diffusive transport around such defects is covered by the Drude model, the ballistic scattering regime was first described by Rolf Landauer, who showed that it leads to residual resistivity dipoles which impose fundamental limits on charge transport \cite{Landauer1957}.

Scanning tunneling potentiometry (STP) \cite{muralt1986scanning} allows to image the electric potential around resistivity dipoles with great precision, by passing a current through a sample and imaging the resulting voltage drop with a scanning tunneling tip. So far, STP has been used to study resistivity dipoles in metallic thin films \cite{muralt1986scanning,Briner1996, Feenstra1998}, reconstructed semiconductor surfaces \cite{bannani2008local, Luepke2015, LuepkeThesis}, graphene \cite{ji2012atomic,Willke2015spatial, willke2017magnetotransport, momeni2018minimum, mogi2019ultimate, sinterhauf2021unraveling, krebs2023imaging, krebs2024imaging,markovic2025intermediate}, and topological insulators \cite{bauer2016nanoscale,Luepke2017}. However, it has become clear that the observed dipoles around {point-like} defects often cannot be unambiguously assigned to either the diffusive or ballistic transport regimes. Possible reasons for this are defect shapes that are not captured by textbook theory, or defect sizes that are 
close to the transition region between diffusive and ballistic transport. 
While there are recent theoretical efforts to describe transport in this transition region \cite{geng2016unified, morr2016crossover}, {resolving the details of the transition in experiments has remained challenging, due to its non-trivial dependence on the defect shapes \cite{markovic2025intermediate}.}

Here we present a systematic analysis of resistivity dipoles at holes in a thin metallic Bi film, and reveal the transition from diffusive to Landauer resistivity dipoles as the defect size decreases below the carrier mean free path. Our results thus provide nanoscale real-space evidence for the fundamental transport limit postulated by Landauer more than 60 years ago. 

{
\section{Origin of the resistivity dipoles} \label{sec:origin}
}
{
The origin of the dipoles is the scattering of charge carriers at the defect boundary, which leads to an accumulation of charge in front of the defect and a depletion behind it.
This charge accumulation has two consequences: 
Inside the defect the electric field is enhanced, 
while outside the defect a local electric field arises that is opposite to the driving field $E_0$.
Thus, the current $\ve{j}$ around the defect is reduced, leading to an increase in resistance in its vicinity. }

\begin{figure}[tbp]
\begin{center}
	\includegraphics[width=\columnwidth]{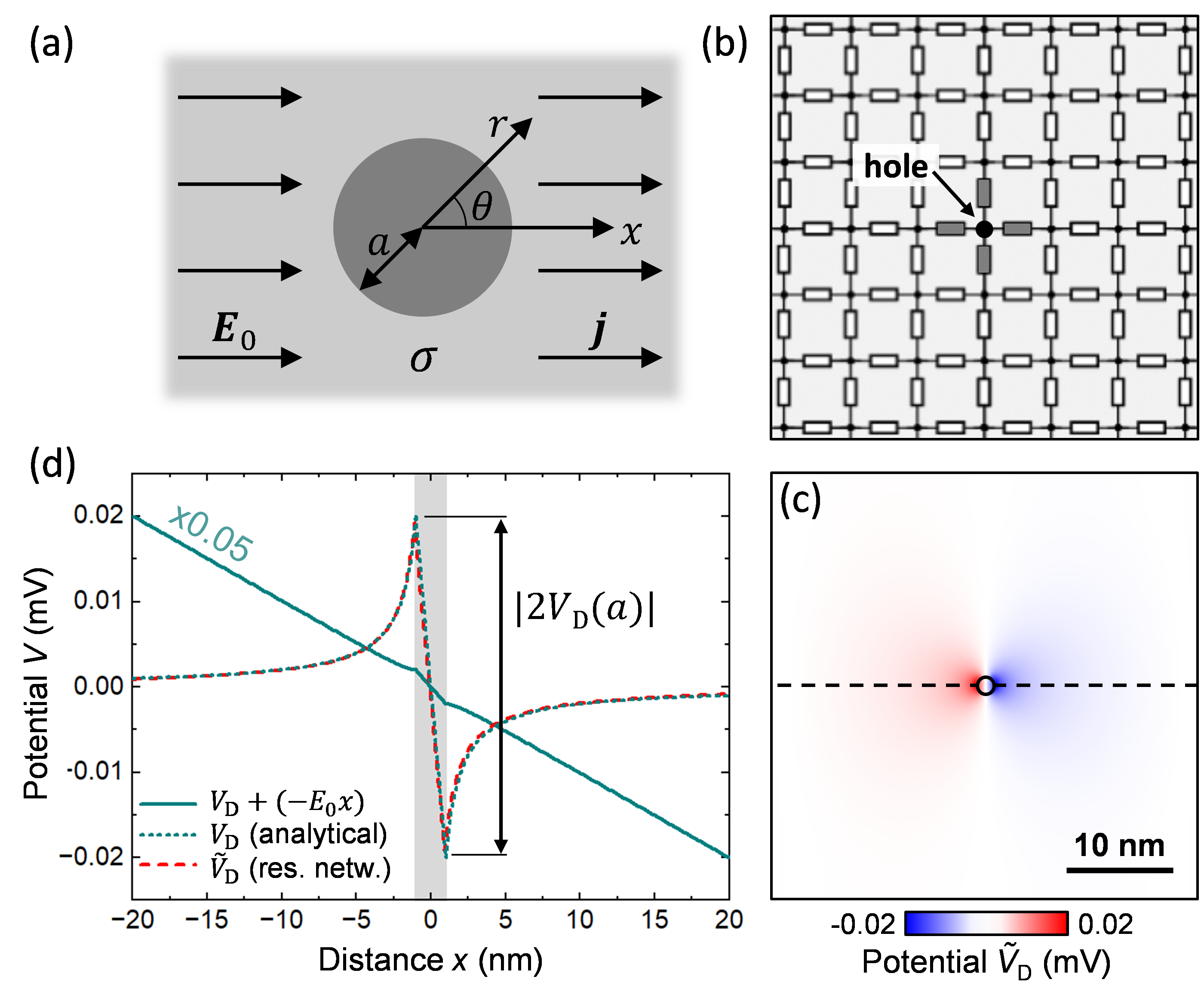}
\end{center}
	\caption{\label{fig:1} \textbf{Resistivity dipole at a point-like defect in a thin film.} (a) Schematic of a circular hole with radius $a$ in a two-dimensional conductor with sheet conductivity $\sigma$. 
    Injecting and draining a homogeneous current density $j$ in lateral (or $x$) direction results in a uniform electric transport field $E_0$ far away from the hole.
    (b) Schematic resistor network, with a black circle in the center representing the hole shown in panel (a). White resistors connect nodes in the film, with resistance according to the film's sheet conductivity $\sigma$, while dark resistors, which connect to nodes inside the insulating hole, block current from entering the hole.
    (c) Dipole potential around the hole (outline indicated by circle) calculated from a $201\times201$ nodes resistor network upon injecting a current in $x$ direction. The linear background corresponding to $E_0=20\rm\, kV/m$ has been subtracted. 
    (d) Cross section of the potential along the dashed line in panel (c) (dashed red curve).
    The hole size is indicated by the shaded region.
    The dotted teal curve shows the analytical solution in the diffusive limit [Eqs.~(\ref{eq1:potentialDipole}) and (\ref{eq2:dipoleMoment_diffusive})]. The solid teal curve shows the total potential before the subtraction of the linear background due to the driving field $E_0$ (scaled by a factor 0.05).
    {The diffusive dipole amplitude at the hole edge is $|2V_\mathrm{D}(a)|$.}
    }
\end{figure}

The scattering of charge carriers at a circular defect (such as a hole) of radius $a$ in a film of sheet conductivity $\sigma$ (modeled as a two-dimensional electron gas {[2DEG]}), through which a current is driven in the $x$ direction (Fig.~\ref{fig:1}a), results in a resistivity dipole moment $p$. This dipole moment generates a dipole potential \cite{Jackson2021,Feenstra1998}
\begin{equation}
    \label{eq1:potentialDipole}
V_{\rm dipole} (r, \theta) = - \frac{p\cos \theta}{r}, \quad \quad r\geq a  
\end{equation}
in the film around the defect, in addition to the externally applied potential $V(r, \theta)=-E_0 x$ that drives the current in the first place. Here, $\theta$ is the angle with respect to the current direction and $r$ is the distance to the center of the defect. In the diffusive limit $a\gg\lambda$, where $\lambda$ is the mean free path of the carriers, the resistivity dipole moment $p$ can be calculated by solving the Laplace equation $\Delta V(r, \theta)=-\nabla \cdot \ve{j}/\sigma=0$ for the potential in the film in which the current density $\ve{j}(r,\theta)=\sigma \ve{E}(r,\theta)$ spreads out.
{Solving the Laplace equation
for a circular hole,
the following boundary conditions apply \cite{CumaThesis}:
Since there are no sources and drains in the area of interest (the current is injected and extracted far away from the defect), the normal component of the current density at the defect boundary must be zero, while the potential (and thus also its tangential gradient $\partial V/\partial \theta$) must be continuous.} 
{It follows the resistivity dipole moment in the diffusive limit} \cite{Jackson2021}
\begin{equation}
\label{eq2:dipoleMoment_diffusive}
    p_{\rm D} = E_0 a^2,
\end{equation}      
\textit{i.e.}, it scales linearly with the area of the defect and the driving electric field. 
As Figs.~\ref{fig:1}b-d show, the dipole potential around a defect in the diffusive regime can be modelled by a resistor network \cite{Luepke2017,LuepkeThesis,CumaThesis}, in excellent agreement with the analytical solution given by Eqs.~(\ref{eq1:potentialDipole}) and (\ref{eq2:dipoleMoment_diffusive}).  As defects become smaller, the diffusive resistivity dipole moment is expected to decrease; 
however, once the defect size approaches the carrier mean free path, the model of diffusive transport reaches its limits. Finally, when $a\ll\lambda$, the defect interacts with ballistically (rather than diffusively) moving carriers. 

{
In this limit, the quasi-Fermi level (electrochemical potential) of right-moving carriers changes sharply (almost discontinuously) across the defect \cite{Datta1997}, which acts as a point scatterer. Since the quasi-Fermi level is a measure of the number of carriers per unit area (carrier density $n$), the latter also drops sharply at the scatterer, creating an electrostatic potential across the defect that (within the defect) adds to the initial driving field. Due to the finite screening length $\xi\gg a$ in the two-dimensional electron gas outside the defect, the accumulated (depleted) charge in front of (behind) the defect extends over a length $\xi$, creating the famous Landauer dipole \cite{Landauer1957}.  Following this argument, it is clear that for ballistic carriers and their scattering the {{longitudonal} defect size} $a_\parallel\ll\lambda$ is not a relevant parameter for the emergence of the Landauer dipole $p_\mathrm{L}$. To estimate its size, we observe that it must {rather} scale with the scattering cross section, which can be approximated by the transverse linear extension $a_\perp$ of the defect. 
{{Therefore,} 
{
the {longitudonal} (in current direction, $a_\parallel$) and transverse (perpendicular to current direction, $a_\perp$) extensions of the defect, 
both play physically distinct roles, even if the defect is (nearly) circular, $a_\parallel\simeq a_\perp$.} 
The resistivity dipole {moment} $p_\mathrm{L}$ should to lowest order also scale linearly with the driving field $E_0$, \textit{i.e.}, $p_\mathrm{L}=\gamma a_\perp E_0$. The different scaling behavior of diffusive and Landauer dipoles with defect size $a=\sqrt{a_\parallel a_\perp}$ implies that at a certain threshold size $a_0$, the function $p_\mathrm{D}(a)$ must cross $p_\mathrm{L}(a)$, such that $p_\mathrm{D}(a)<p_\mathrm{L}(a)$ for $a_\parallel<a_{0,\parallel}$. We can expect that $a_{0,\parallel}\sim\lambda$. Equating $p_\mathrm{L}(a_0)=\gamma a_{0,\perp} E_0\overset{!}{=}p_\mathrm{D}(a_0)=a_{0,\parallel} a_{0,\perp} E_0$, we find $\gamma=a_{0,\parallel}$ and thus $p_\mathrm{L}(a)=a_{0,\parallel}a_\perp E_0$ (for general $a,\,a_\perp$). Inserting $E_0=j/\sigma$ and using the conductivity $\sigma=ne^2\lambda/(\hbar k_\mathrm{F})$ and the carrier density $n=k^2_\mathrm{F}/(2\pi)$ of a two-dimensional electron gas, we obtain $p_\mathrm{L}= \frac{2\pi a_{0,\parallel}}{\lambda}\frac{\hbar j}{k_{\rm F}e^2} a_\perp$, which apart from the unknown factor $a_{0,\parallel}$ resembles Eq.~(\ref{eq3:dipoleMoment_Landauer}). 
A detailed analysis \cite{Sorbello1988, markovic2025intermediate} reveals that  {$a_{0,\parallel}=8\lambda/(3\pi)$}, finally yielding
}
\begin{equation}
    \label{eq3:dipoleMoment_Landauer}
    p_{\rm L} = \frac{16}{3}\frac{\hbar j}{k_{\rm F}e^2} a_\perp, 
\end{equation}
where $k_{\rm F}$ is the Fermi wave vector.
} 

{Using $j=\sigma E_0$, we can rewrite Eq.~(\ref{eq2:dipoleMoment_diffusive}) as $p_\mathrm{D}= \sigma^{-1}a_\parallel a_\perp  j$, where $\sigma^{-1}$ is the sheet resistivity of the defect-free film. Comparing to Eq.~(\ref{eq3:dipoleMoment_Landauer}), we can define a resistivity dipole $ \rho_\mathrm{dipole}$ with which both the diffusive resistivity dipole moment $p_\mathrm{D}$ [Eq.~(\ref{eq2:dipoleMoment_diffusive})] and the Landauer (ballistic) resistivity dipole moment $p_\mathrm{L}$ [Eq.~(\ref{eq3:dipoleMoment_Landauer})] can be expressed in the same generic form  
{
\begin{align}  
    p_\mathrm{D/L}=\rho^\mathrm{D/L}_\mathrm{dipole}a_\perp j, 
\end{align}
where
\begin{align}
    \rho_\mathrm{dipole}^\mathrm{D}=\sigma^{-1}a_\parallel
\end{align}
and
\begin{align}
    {\rho^\mathrm{L}_\mathrm{dipole}=\frac{8\lambda}{3\pi\sigma}=\frac{16\hbar}{3k_\mathrm{F}e^2}}. \label{eq:rho_L}
\end{align}
}We note that $\rho_\mathrm{dipole}^\mathrm{D}$ scales with the {longitudonal} extension $a_\parallel$ of the defect, while $\rho_\mathrm{dipole}^\mathrm{L}=(8/3\pi)\sigma^{-1}\lambda$ depends on the mean free path $\lambda$ instead. Both resistivity dipoles are proportional to the sheet resistivity $\sigma^{-1}$.}

Analyzing the dipole potential given by Eq.~(\ref{eq1:potentialDipole}) at the edges of defects ($r=a$) parallel to the current direction ($\theta=0^\circ)$, distinct dependencies on the defect size follow from Eqs.~(\ref{eq2:dipoleMoment_diffusive}) and (\ref{eq3:dipoleMoment_Landauer}):
$V_\mathrm{D} \propto a$, but $V_\mathrm{L} = \mathrm{const.}$ (Ref. \onlinecite{Feenstra1998}). Thus, it is possible to observe {dipoles in} the different transport regimes in one and the same sample, if it contains defects of sufficiently different sizes. 

\begin{figure}[tbp]
\begin{center}
\includegraphics[width=\columnwidth]{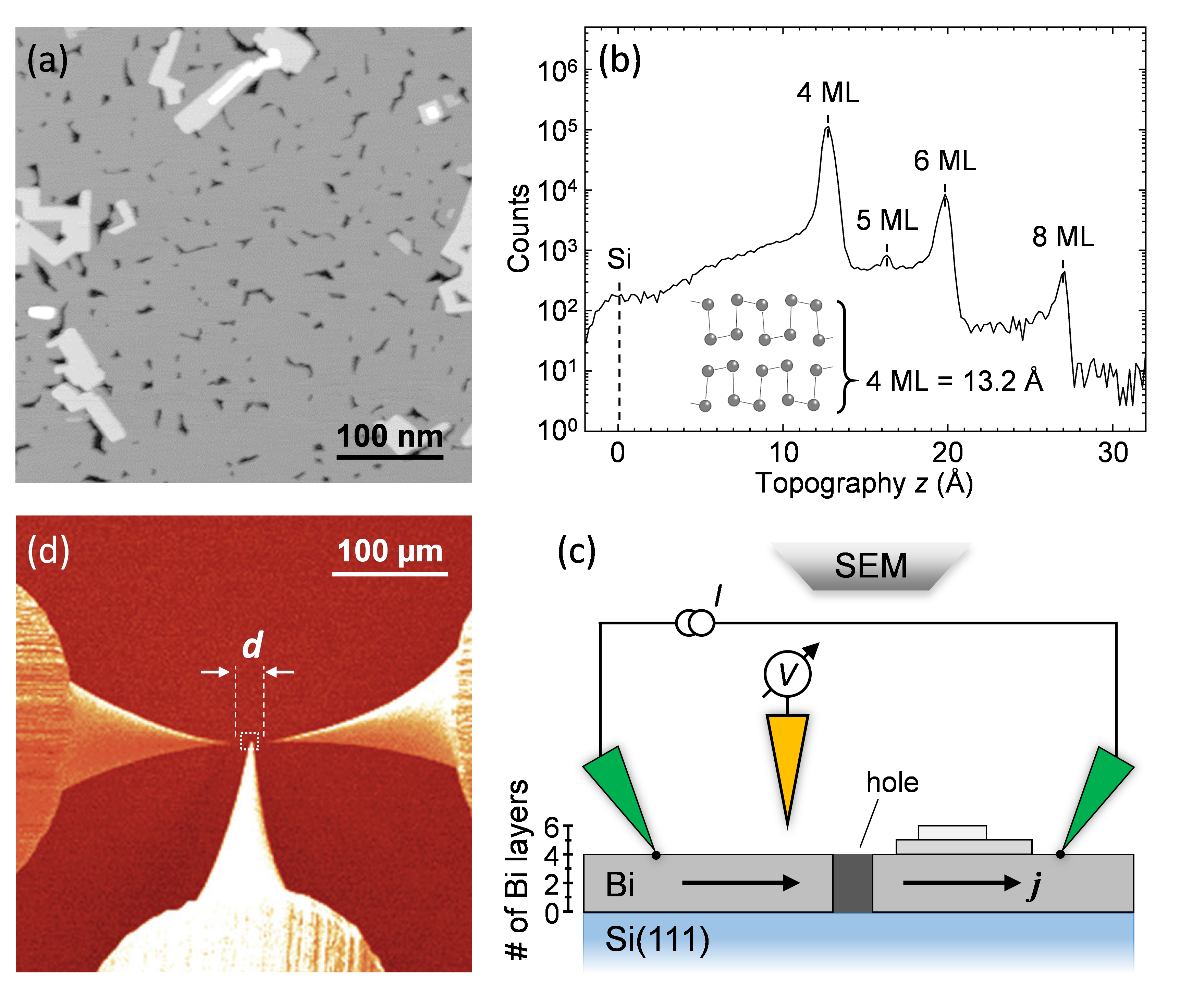}
\end{center}
\caption{\label{fig2:STP_Bi} \textbf{Scanning tunneling potentiometry on thin Bi films.} (a) Scanning tunneling micrograph of 4\,ML Bi deposited on a Si(111)-$7\times 7$ sample. 
(b) Histogram of panel (a), showing that the majority of the Bi film has a thickness of 4\,ML (inset) with holes extending down to the Si substrate [dark areas in panel (a)] and a small number of additional islands on top (bright areas). 
(c) SEM image of the tips during the potentiometry experiments. The distance between the current-injecting tips is $d=25\,$\textmu m.
(d) Schematic of the sample cross section and potentiometry setup implemented into a multi-tip STM. The two outer tips (in point contact) inject a lateral current density $j$ into the film, while the central tip (in tunneling mode) maps the electrochemical potential. The tips are positioned under scanning electron microscope (SEM) observation. 
}
\end{figure}

{
\section{Methods}
\subsection{Sample preparation}
}
{We realized the above described} scenario by depositing four monolayers (4\,ML) of Bi on a Si(111)-$7 \times 7$ substrate. 
{For this purpose, a}{n 8 $\times$ 4 mm$^2$ piece of Si(111) (miscut $<0.15^\circ$, $\rho_{\rm Si} \approx 700\rm\, \Omega cm$) was degassed under ultra-high vacuum (UHV) conditions at $T_{\rm S} \approx 700^\circ\rm C$ for several hours, followed by repeated flash-annealing cycles at $1230^\circ\rm C$ for 30\,s each. After flashing, the sample was first quenched to $T_{\rm S} = 1050^\circ\rm C$, followed by cooling to $T_{\rm S} = 950^\circ\rm C$ at a relatively slow rate of $\sim 1 \,{\rm K}/\mathrm{s}$. The sample was then quenched a second time, this time to $T_{\rm S} = 850^\circ\rm C$ and held at this temperature for 30\,min to form a well ordered Si(111)-$7\times 7$ surface structure, followed by a final quench to room temperature. During sample preparation, the direction of the heating current (technical current direction) was in the "step-up" direction, so that the resulting Si(111)-$7\times 7$ surface showed parallel step bunches with several hundred nanometer wide flat terraces in between \cite{GibbonsThesis, RomanyukThesis}. 4\,ML of Bi were deposited on the Si(111)-$7\times 7$ surface at room temperature from a Knudsen effusion cell at a deposition rate of $1\rm\,ML/min$, where $1\rm\,ML=9.28 \times 10^{14}\rm\, atoms/cm^2$.
The resulting Bi films have a (012) surface orientation and are in the black-phosphorus phase \cite{Nagao2005}{, and} show large atomically flat terraces (gray in Fig.~\ref{fig2:STP_Bi}a) with naturally occurring holes (dark) and a small number of adislands (bright) \cite{Nagao2005, Yaginuma2007}. 
A histogram of the STM topography (Fig.~\ref{fig2:STP_Bi}b) shows a dominant peak at 4\,ML and less pronounced peaks for 5, 6, and 8\,ML above the background; the latter stems from the step edges of the islands and the holes. The holes are irregularly shaped and their sizes vary from $\sim 1$ to 50 nm (Fig.~\ref{fig2:STP_Bi}a). 
}
\begin{figure*}[htbp]
\begin{center}
	\includegraphics[width=2\columnwidth]{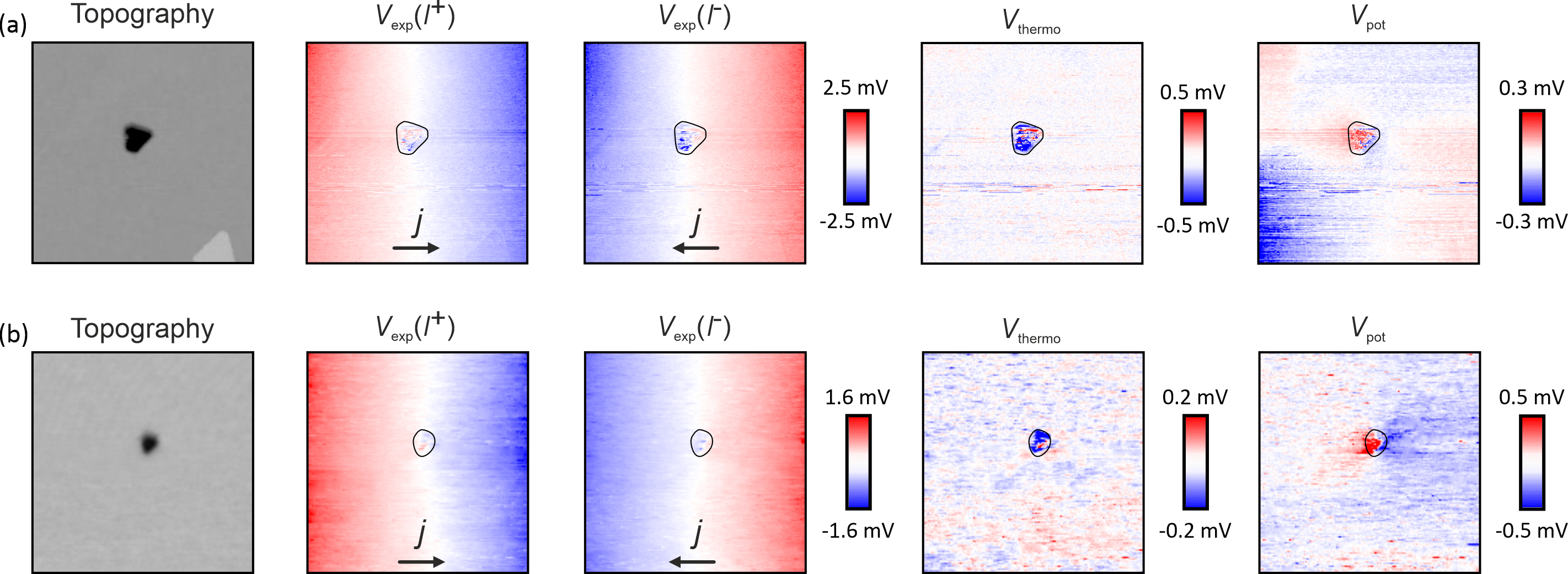}
 \end{center}
	\caption{\label{fig:Sup1}
    {
    \textbf{Raw potential maps for opposite current directions and resulting thermovoltage and potential maps.}
     (a) Topography, potential maps resulting from opposite current directions before linear background subtraction and calculated thermovoltage $V_\mathrm{thermo} = (V_\mathrm{exp}(I^{+})\  +\  V_\mathrm{exp}(I^{-}))/2$, and potential maps $V_\mathrm{pot} = (V_\mathrm{exp}(I^{+})\  -\  V_\mathrm{exp}(I^{-}))/2$ for the hole shown in Fig.~\ref{fig:Sup2}c.
     (b) Same as panel (a) but for Fig.~\ref{fig:Sup2}d.
    { The current densities for panels (a) and (b) are $j=(7.90 \pm 0.56)\rm\,A/m$ and $(7.41 \pm 0.23)\rm\,A/m$, respectively.}
    }}
\end{figure*}
\begin{figure*}[bt]
\begin{center}
 	\includegraphics[width=2\columnwidth]{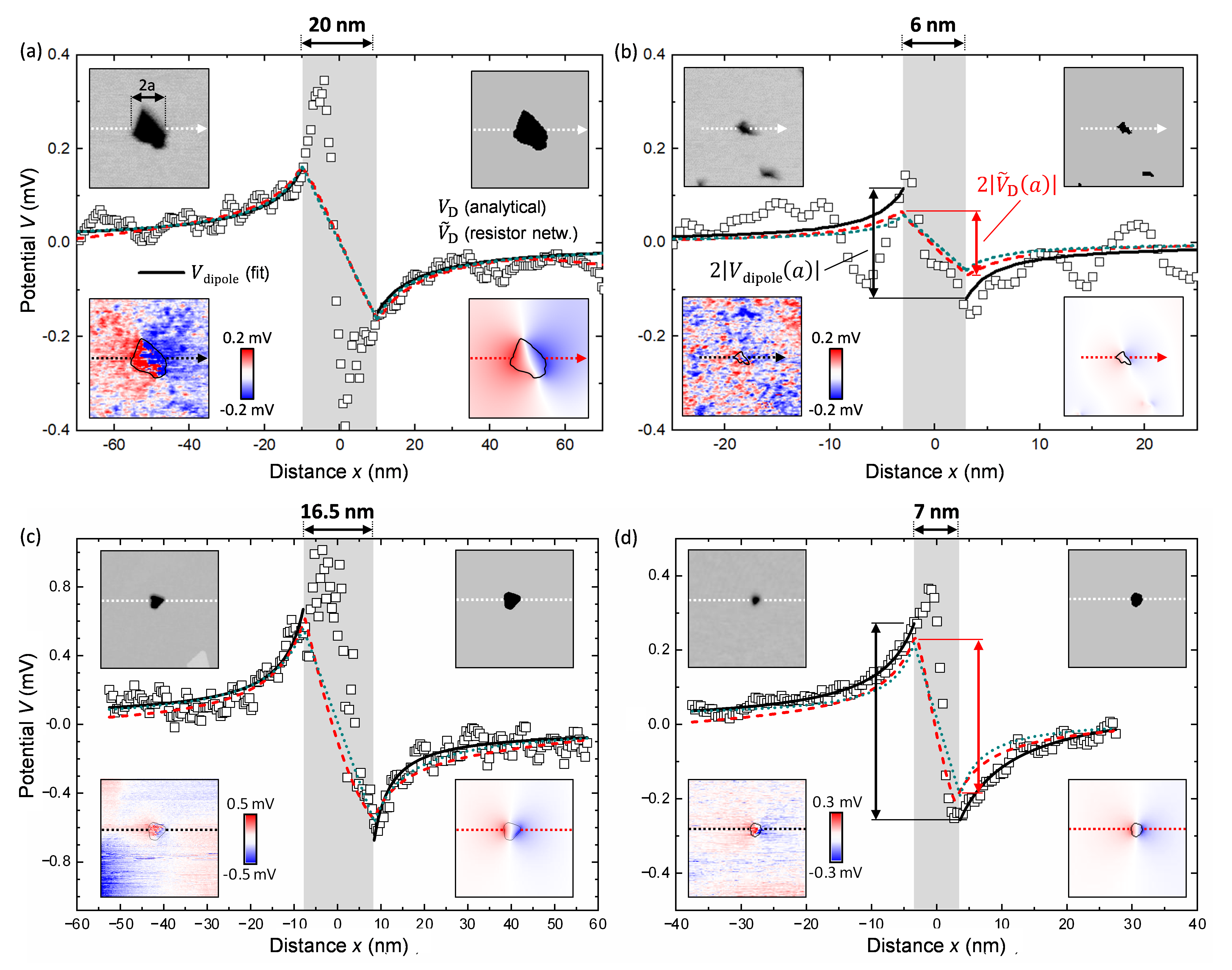}
  \end{center}
 	\caption{\label{fig:Sup2}
     {\textbf{Exemplary resistivity dipoles}.
     (a, b) Cross sections of the potentials along the dotted lines in the insets, recorded at two exemplary holes from data set A. 
     The hole sizes are $2a\simeq {20}$ and ${6}\,\mathrm{nm}$, respectively.
     The potential data was recorded without current reversal. As a result, the dipole potentials are superimposed with thermovoltage signals. Insets: Experimental topography (top left) and potentiometry map (bottom left), hole map resulting from threshold detection (top right), and potential map resulting from resistor network calculation (bottom right). 
    All maps in the panel cover the same $x$ range as the main plot.
    The shaded region in the main panel corresponds to the cross section of the {hole in the upper right inset}. 
    (c, d) Same as panel (a,b) but for two dipoles from data set B, recorded with the current-reversal scheme. As a result, thermovoltages are excluded, resulting in clearer transport dipole signatures. 
    The holes sizes $2a\simeq {16.5}$ and ${7}\,\mathrm{nm}$ are comparable to those in panels (a,b). 
     {The overshoot of the potentials inside the holes is likely caused by a tip artifact (see section \ref{sec:tip_shape}).
     Fits to the data outside of the holes} for data set A were performed with the strict fit function Eq.~(1). {Fits for data set B were performed with the relaxed fit function Eq.~(\ref{eq:relaxed_fit}). 
     For the large holes, the measured dipole potential (black squares) and fits following Eq.~(\ref{eq1:potentialDipole}) to the experimental data 
    {(solid black curves)}
    agree well with the calculated analytical dipole potential $V_\mathrm{D}$ in the diffusive limit [Eq.~(\ref{eq2:dipoleMoment_diffusive}), dotted teal curve], and with the diffusive dipole potential resulting from the resistor network calculation $\tilde V_\mathrm{D}$ (dashed red curve).
    Around the small holes, $V_\mathrm{dipole}$ is significantly larger than both $V_\mathrm{D}$ and $\tilde V_\mathrm{D}$.  
    The current densities are (a) $j=(3.58 \pm 0.36)\rm\,A/m$, (b) $(4.33 \pm 0.43)\rm\,A/m$, (c) $(7.90 \pm 0.56)\rm\,A/m$, and (d) $(7.41 \pm 0.23)\rm\,A/m$, respectively.}}
     }
\end{figure*}
\begin{figure*}[htbp]
\begin{center}
	\includegraphics[width=2\columnwidth]{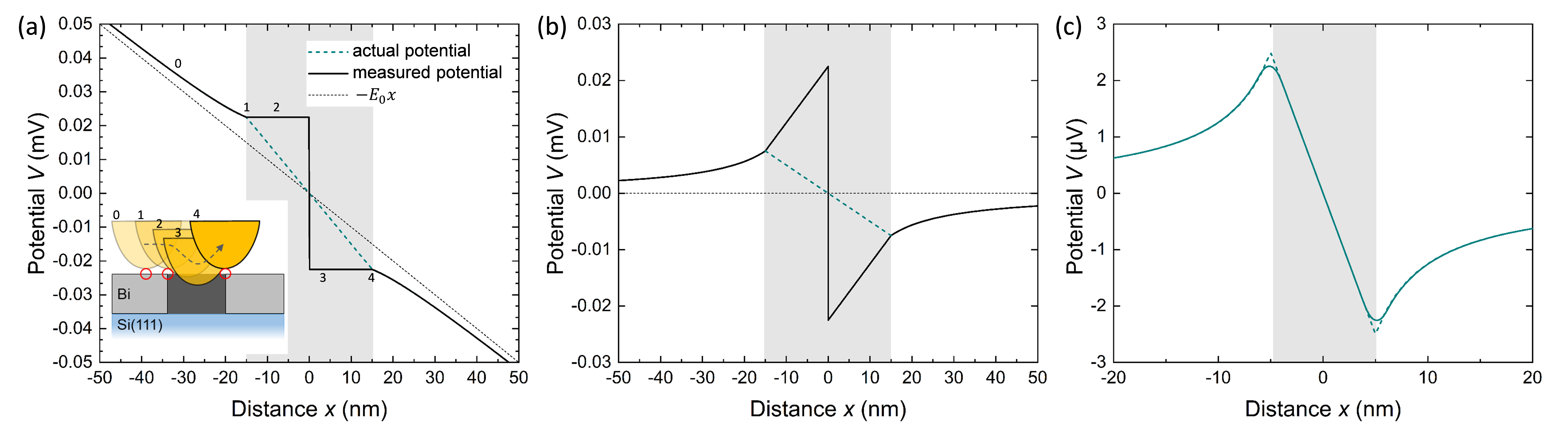}
	\end{center}
 \caption{\label{fig:Sup3}
    {
    \textbf{Simulation of the measured versus the actual potential across a hole.}
     (a) Inset: Schematic of a blunt tunneling tip scanning across a hole in constant-current mode. Several distinct tip positions are labeled with numbers 0 to 4. The tip positions from the inset are also indicated in the main graph. The gray-shaded area marks the size of the hole {($a=15\rm\,nm$)}. Due to the bluntness of the tip compared to the step edge, the measured potential deviates from the actual potential.
    (b) Measured (solid black line) and actual (teal dotted line) potential after subtracting the linear background $-E_0x$.
    (After Ref. \onlinecite{CumaThesis}, but we note that in Fig.~3.10 of that reference, the situation for a defect with increased conductivity is shown.)
    {(c) Theoretical (dotted line) and simulated broadened dipole potential (solid line) resulting from a moving average filter with $2\rm\,nm$ window size, illustrating the reduction of the potential near the hole edge ($a=5\rm\,nm$).}
    }}
\end{figure*}
\begin{figure}[htbp]
\begin{center}
 	\includegraphics[width=0.8\columnwidth]{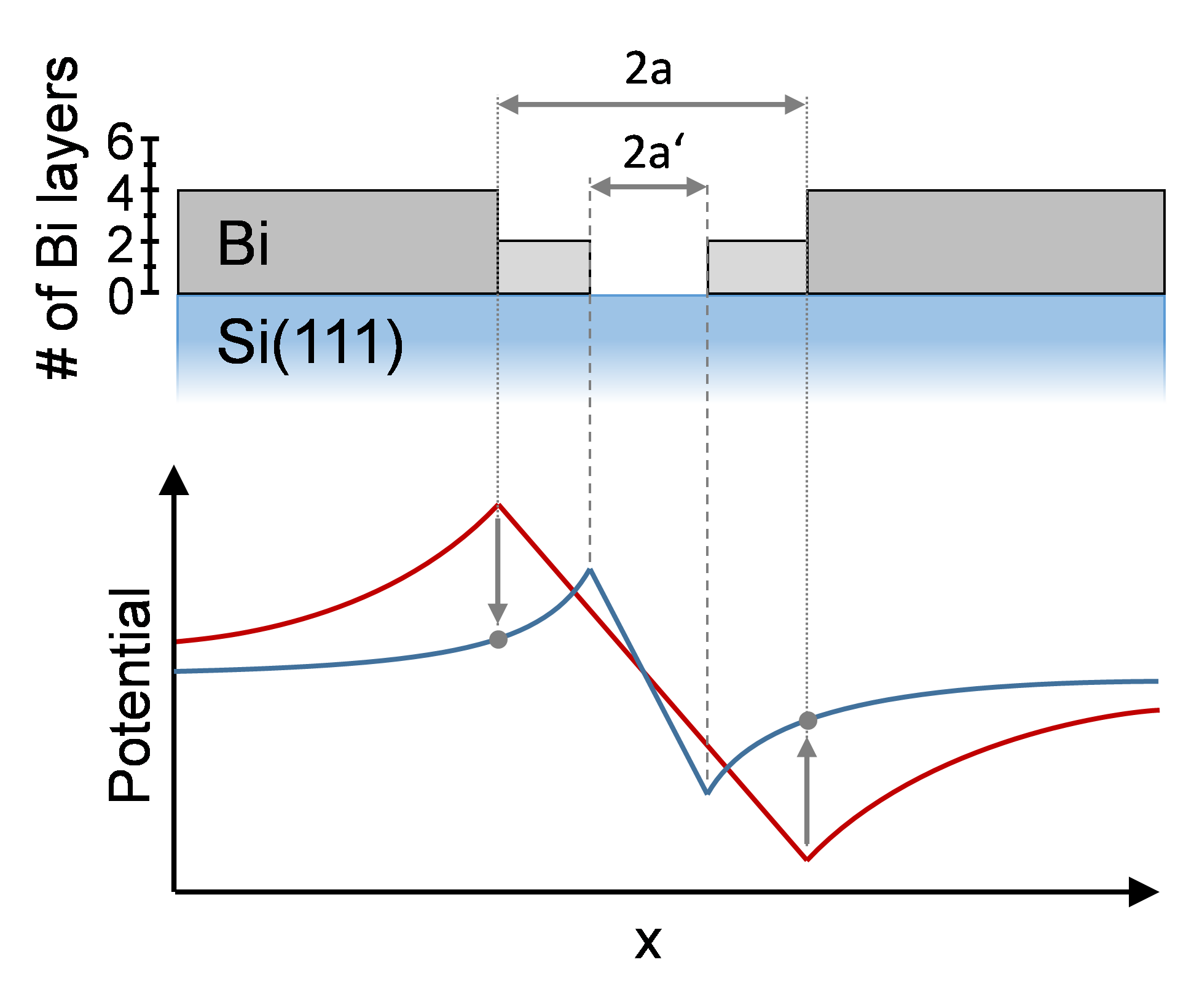}
  \end{center}
 	\caption{\label{fig:Sup4}
      {
     \textbf{Influence of additional subsurface Bi layers in the hole on the measured dipole potential.} The schematic displays a hole and two associated dipole potentials. The threshold detection scheme detects the edge of the 4\,ML film (dark gray), corresponding to a hole size $2a$. As a consequence, the potential is evaluated at the edge of the 4\,ML film. If no subsurface layers were present inside the hole, this would yield the extrema of the red potential profile as the relevant dipole potential. However, if Bi layers extend into the hole (\textit{e.g.}, the 2\,ML film displayed in light gray), then the extrema of the dipole potential will be located further into the (large) hole, where the additional, thinner layers terminate (corresponding to a hole size $2a'$). Because the hole in the 2\,ML film is smaller than the one in the 4\,ML film, its resulting dipole potential (blue curve) also has a smaller maximum amplitude. In addition, the potential evaluated at the edge of the 4 ML film (corresponding to size 2a) does not correspond to the extrema of the blue dipole potential. Thus, the \textit{experimentally detected} dipole at the \textit{experimentally detected} hole edges (indicated by the gray dots) is not only smaller than it would be if no layers were present inside the hole (indicated by the arrows), but also smaller than the actual dipole of the $2a'$ hole. We note that in this schematic, we assumed the conductivity of the 2\,ML film to be the same as that of the 4\,ML film. Realistically, the conductivity of the 2\,ML film is lower, which would result in a superposition of two transport dipoles. However, the resistivity dipole measured at the threshold-detected edge would still be smaller compared to the case if no film extended into the hole.
    }}
\end{figure}

{
\subsection{Scanning tunneling potentiometry}
\label{sec:STP} 
We performed STP measurements {on two separately prepared samples (data sets A and B)} using two home-built multi-tip scanning tunneling microscopes \cite{cherepanov2012ultra, Luepke2015, voigtlander2018invited, LuepkeThesis} (see Fig.~\ref{fig2:STP_Bi}c,d), which allow us to analyze the freshly prepared samples under ultra-high vacuum conditions. 
This is critical because standard lithography and transport methods are not compatible with the air-sensitive Bi films.
{Both multi-tip STMs were} operated at room temperature and equipped with electrochemically etched tungsten tips cleaned \textit{in situ} by resistive heating and sputtering. The pressure during the measurements was $p \le 2 \times 10^{-10}$ mbar and we did not observe any surface contamination even a few weeks after the measurements. }

{The STP measurements were performed on flat terraces away from {substrate} step bunches, where the flow of a lateral current is not influenced by any substrate steps. The sample surface was contacted with two tips to inject a lateral current $I$ in the range $0.1$\,mA~to~$1$\,mA at a tip-to-tip distance of $d=5\,$\textmu m~to~$25\,$\textmu m. At these tip distances, no significant current flows through the Si substrate, as the two-point resistance of the used Si substrate is much larger than that of the 4\,ML Bi film. }

{The topography and the electrochemical potential in the central region between the two current-injecting tips were simultaneously measured by scanning a third tip in tunneling contact, using an `interrupted feedback' technique \cite{Luepke2015, druga2010versatile}: The sample was scanned with constant-current feedback to determine the sample topography. At every image pixel, however, the constant-current topography feedback was switched off temporarily and potential feedback was switched on. For potential feedback, the tip was fixed at its present height, while the voltage applied to the tip was adjusted by a software feedback loop until the tunneling current was nulled. The voltage at which this occurred was then recorded as the local potential of the sample at this pixel. The initial tip bias was then restored and the scan continued to the next pixel. The resulting spatial and potential resolution was $\sim1\,$\AA\ and $\sim5\,$\textmu V for typical tunneling parameters $V_{\rm tip} = -5\rm\,mV$, $I_\mathrm{t} = 10\rm\, pA$. For {further} details on the implementation of this technique, see Refs. \cite{Luepke2015, LuepkeThesis}.}

{ The current density in the center between the two current-injecting tips is given by $j = 2I/(\pi d)$ \cite{ji2012atomic}. 
However, tip positioning errors can lead to significant deviations between the actual local current density at the hole and the one calculated by the above equation.
To determine the current density at the investigated holes more accurately, we therefore employed the linear background slope $E_0$ and used the film conductivity $\sigma=0.22\,\mathrm{mS}/\square$ determined from four-point measurements \cite{CumaThesis} to calculate the local current density as $j=E_0\sigma$ (the {resulting} current densities are quoted in the respective figure captions below)}.

{
\subsection{Thermovoltage}
\label{sec:thermovoltage}
Thermovoltages, which arise from small temperature differences between tip and sample {[}$V_\mathrm{thermo}\sim(T_\mathrm{tip}^2-T_\mathrm{sample}^2)${]}, are known to be present in scanning tunneling potentiometry \cite{Luepke2015}. Such thermovoltages are superimposed on the voltage drop resulting from the injected current and can lead to a nonuniform background, which complicates the extraction of the transport field $E_0$ and the resistivity dipoles. In detail, the measured potential signal $V_\mathrm{exp} = V_\mathrm{pot} + V_\mathrm{thermo}$ is composed of the sought-after potential-derived signal $V_\mathrm{pot}$ and the (in our case unwanted) thermovoltage signal $V_\mathrm{thermo}$.
}

{To exclude thermovoltages from the measured potentials, we implemented a dual-pass measurement scheme: Each scan line of the potentiometry scan was scanned twice, once with forward current direction ($I^{+}$) and a second time with the inverted current direction ($I^{-}$). While the thermovoltage is independent of the current and its direction, the potential-derived  contribution to the measured signal inverts with the current direction,\ $V_\mathrm{pot}(I^{-}) = - V_\mathrm{pot}(I^{+})$. 
Thus, the thermovoltage component to the measured signal can be removed by subtracting the measured potential signals for the two opposite current directions: $V_\mathrm{pot} = (V_\mathrm{exp}(I^{+})\  –\  V_\mathrm{exp}(I^{-}))/2$. The resulting $V_\mathrm{pot}$ thus intrinsically reflects the transport dipoles from \textit{both} current directions. 
Correspondingly, the thermovoltage is calculated as $V_\mathrm{thermo} = (V_\mathrm{exp}(I^{+})\  +\  V_\mathrm{exp}(I^{-}))/2$. In Fig.~\ref{fig:Sup1}, the potential data for the two opposite current directions, recorded for two holes from data set B, are shown together with the resulting $V_\mathrm{pot}$ and $V_\mathrm{thermo}$. To record this data, we implemented a current source which allows a current {direction} reversal while maintaining the same absolute current.}

{In measurements belonging to data set A, the dual-pass measurement scheme was not yet used and accordingly no correction for thermovoltages was applied. In Fig.~\ref{fig:Sup2}, we compare the potentials extracted from data set B (with thermovoltage correction, as explained in the previous paragraph) with those of data set A, for two pairs of comparable holes, one large and one small. Thermovoltage contributions appear as oscillations in the potential line scans in Fig.~\ref{fig:Sup2}a,b and a grainy potential map in the lower left insets in Fig.~\ref{fig:Sup2}a,b. As apparent in Fig.~\ref{fig:Sup2}c,d, for data set B the thermovoltage correction yields cleaner signatures of the transport dipoles. Comparing the two data sets A and B, however, we note that, in spite of the oscillations, the dipole potential amplitudes $|V_{\rm dipole}(a)|$ resulting from the fits {of Eq.~(\ref{eq1:potentialDipole}) to} the experimental data on both sides of the holes are not significantly affected by the oscillations due to the thermovoltage (see, \textit{e.g.}, Fig.~\ref{fig:Sup2}a,b vs.~c,d). Correspondingly, the transition between transport regimes is {expected to be} found consistently in both data sets, \textit{i.e.}, independent of the removal of the thermovoltages as described above.
}

{
\subsection{Detection of hole size}
\label{sec:detection_hole_size}
To detect the edges of the holes, we used a threshold detection algorithm on the STM topography data. The latter were acquired simultaneously with the potential data (see section~\ref{sec:STP}). Due to the finite tip radius in our experiments (see section~\ref{sec:tip_shape}), the edges of the holes are smeared out in the STM topography, requiring a careful choice of the threshold value $\delta_z$ for different tunneling tips. For the purpose of choosing the correct threshold, we tested our detection procedure on holes that we knew to be in the diffusive regime: We set $\delta_z$ such that for large holes $\gg a_0$ the {dipole amplitudes} are located on the line expected for the diffusive regime {($V_\mathrm{D}\propto a$)}, \textit{i.e.}, the resistor network calculations {(\textit{c.f.} section \ref{sec:res_net})} and experimental data coincide \cite{CumaThesis}. Since we did not change the tip during the recording of a given data set, we used a single threshold value for all holes within each of the two data sets. Specifically, we found $\delta_z =0.3\,$nm for data set A and $\delta_z =0.15\,$nm for data set B. To determine the hole sizes, $\delta_z$ is referenced to the flat terraces surrounding the holes, and a hole is detected once the topography signal falls $\delta_z$ below the value on its surrounding terrace.}

{
\subsection{Resistor network calculations}\label{sec:res_net}
{The analytical dipole $V_\mathrm{D}$ [Eq.~(\ref{eq1:potentialDipole})] is strictly applicable only to circular holes. 
Therefore, t}o accurately account for the influence of the hole shape{s}, we performed resistor network calculations for each hole, which allow us to determine the diffusive dipole potential around arbitrarily shaped defects \cite{Luepke2017, Homoth2009, LuepkeThesis}.
Based on the results of the hole detection algorithm (see Section~\ref{sec:detection_hole_size}), we created a mask of all topography data points below the threshold, which we used as the hole geometry for the resistor network calculations, with which we calculated the diffusive dipole potentials $\tilde V_\mathrm{D}$ {around the experimental holes}.}

{The resistor networks were implemented on a square lattice, with the node potentials given by a vector {$\ve{V}$}. Neighboring nodes are connected by resistors corresponding to either the Bi-film conductivity ($\sigma=0.22\,\mathrm{mS}/\square$, Ref.~\cite{CumaThesis})  on the terraces or the substrate conductivity ($\sigma_\mathrm{Si}=9\,$\textmu S$/\square$, Ref.~\cite{just2015surface}) in the holes. All resistors were entered into a conductivity matrix ${ S}$. 
We then imposed a current between the left and right boundaries of the system and used Kirchhoff's current law to numerically solve the linear system of equations {$\ve V = { S}^{-1}\ve I$} for the potential distribution, where {$\ve{I}$} is the sum of incoming and outgoing currents at each node. The resulting potential distributions were plotted as images in which each pixel corresponds to one node. If the node-to-node distance is significantly smaller than the mean free path $\lambda$ of the charge carriers, such model simulations accurately reproduce diffusive transport around a defect  (Fig.~\ref{fig:1}). Further details can be found in Refs. \onlinecite{LuepkeThesis, CumaThesis}.}

{
\subsection{Influence of the tip shape and lateral potential broadening}
\label{sec:tip_shape}
A finite tip radius (expected to be large compared to the smallest holes) { does not only lead to a smoothing of the edges in the STM topography, but also to an apparent extension of the dipole potential into the holes, because the atomically sharp hole edges make the tunneling channel move across the surface of the blunt tip.} 
{The mechanism} is schematically explained in Fig.~\ref{fig:Sup3}.}

{In detail, the tunneling setpoint current during the topographic constant-current scan maintains a constant tip-sample distance, corresponding to the radius of the red circles in the inset of Fig.~\ref{fig:Sup3}a.
Outside the hole (tip position 0), the tunneling contact always occurs between the tip apex and the flat surface (left red circle).
However, once the tip apex is in the hole (tip positions 1, 2), the tunneling current still flows from the left edge of the hole (middle red circle), but to different positions on the tip surface. This is an immediate consequence of the blunt tip shape.
Beyond the half-way point across the hole, the tunneling contact jumps to the opposite edge of the hole (tip position 3 and right red circle), where it remains until the tip is outside the hole again (tip position 4).
Since in the potential feedback mode used during the potentiometry measurement (see Section \ref{sec:STP}) the bias voltage is adjusted until the measured tunneling current is nulled, the measured potential supposedly measured \textit{within} the hole is thus determined by the tip shape and the potential \textit{at the hole edges}:
As the tip enters the hole from the left, a constant potential corresponding to the potential at the left hole edge is measured for a certain range of $x$ values into the hole, until a sudden jump to a second constant value corresponding to the potential at the right hole edge occurs, which in turn continues until the tip exits the hole.
After subtracting the linear background $-E_0x$ (Fig.~\ref{fig:Sup3}b), the measured potential inside the hole shows an {overshoot and deviates from} the actual potential in the hole (dashed blue line). 
Toward the center of the hole, the overshoot is disrupted by a sharp step once the tunneling contact jumps from the left to the right hole edge. Thus, to exclude the effect of finite tip radii, we always extracted $V_\mathrm{dipole}$ from the fits to the data on the flat terraces outside the holes.}

{Even in cases when the overshoot discussed in the previous paragraph is not {present}, we still may encounter artifacts in the measured potential close to the hole edges. Experimental noise in the $x, y$ tip position (such as due to external mechanical vibrations or electronic noise in the $x$ and $y$ piezo voltages) and signal post-processing can lead to a spatial broadening of the measured potential. This especially affects the dipole potential near the hole edges and leads to a lowering of the potential just outside of the holes, as displayed schematically in Fig.~\ref{fig:Sup3}c. {As a result of this, the experimental data underestimates the theoretical dipole Eq.~(\ref{eq1:potentialDipole}) near the hole edges}. 
}

{In this context, we also evaluated the systematic error of the extracted potential due to an overestimation of the hole size. Such an overestimation could occur in the situation depicted in Fig.~\ref{fig:Sup4}, where 2\,ML Bi partially fill the bottom of a larger hole, thus forming a smaller hole with diameter $2a'$. In this case, the `edge' potential of the smaller hole will not be sampled at its real edges (diameter $2a'$), but at the edges around a  somewhat larger diameter $2a$, where the threshold detection algorithm will detect the hole edge. Clearly, at this position the modulus of the real (\textit{i.e.}, smaller) hole's potential (blue line) will be \textit{lower} than its maximum value at the actual hole edges around the diameter $2a'$, because outside the hole the potential approaches zero. Since in the diffusive regime a hypothetical hole of larger diameter $2a$ will anyway have a larger peak  potential (red curve) than the smaller hole (blue curve), the measured potential of the smaller hole (indicated with grey dots in Fig.~\ref{fig:Sup4}) will be substantially smaller than the peak potential of the hypothetical large hole (grey arrows).    
We can thus conclude that in a situation as shown in Fig.~\ref{fig:Sup4}, the above-mentioned effects imply that the real (small) hole's `edge' potential will be measured at an overestimated radius (corresponding to a hypothetical larger hole), which systematically and significantly  underestimates the correct edge potentials of both of the real (small) and the hypothetical (large) holes. To summarize the above considerations, a partial coverage of the bottom of a hole with a thinner Bi film cannot explain 
{an \textit{increase} in the dipole amplitude above the diffusive dipole $V_\mathrm{D}\propto a$}
}

\begin{figure*}[bt]
 \begin{center}
 	\includegraphics[width=2\columnwidth]{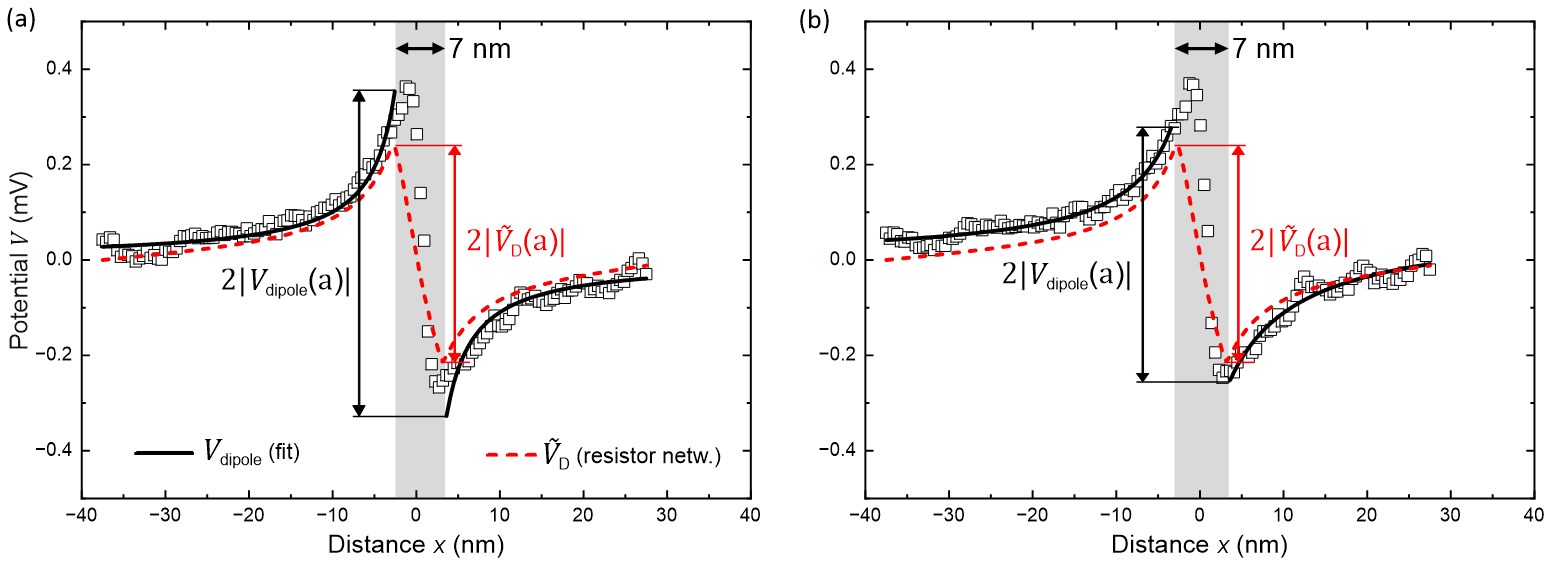}
   \end{center}
 	\caption{\label{fig:Sup5}
    {
     \textbf{Comparison of the relaxed and the strict fit routines.} The experimental data belong to the 7-nm hole of Fig.~\ref{fig:Sup2}d (belonging to data set B), {measured at $j= (7.41 \pm 0.23) \rm\,A/m$}.
     (a) Strict fit using Eq.~(1) simultaneously on both sides of the hole (black line).
     (b) Relaxed fit using Eq.~(\ref{eq:relaxed_fit}) separately and independently on both sides of the hole (black line).
    }}
 \end{figure*}

\section{Results}
Figure \ref{fig:Sup2} shows cross sections of the measured dipole potentials (after linear background subtraction, compare Fig.~\ref{fig:1}d) of two exemplary holes {from each data set}, one large and one small.
The bottom left insets in Fig.~\ref{fig:Sup2} display potential maps around the holes, while the main panels show potential profiles along the dotted lines in the left insets in each panel. 
{The result of the threshold-detection algorithm (see section~\ref{sec:STP}) is shown in the top left insets and resulting hole maps are shown in the top right insets in Fig.~\ref{fig:Sup2}.}
We identified the hole size $2a$ (shaded in gray in the main panels) as the extension of the holes along the white dotted cross sections in the top insets. 

{
\subsection{Fitting the potential data} \label{sec:fitting}
}
{
To quantify the measured dipole potentials $V_\mathrm{dipole}(\pm a)$ at the hole boundaries in front of and behind the hole (with respect to the current direction), we employed two different fitting approaches. First, the  potential profile through the hole was fitted with Eq.~(1). Notably, experimental data points in the hole (where the topography data of the cross section lies below the threshold) were excluded to avoid tip artifacts (see section~\ref{sec:tip_shape}). We found that fitting Eq.~(1)  simultaneously to the experimental data on both sides of a hole in some cases led to 
deviations of the fitted curve from the experimental data points. {This can have several reasons:  One cause could be the overshoot sketched in Fig.~\ref{fig:Sup3} and discussed in section~\ref{sec:tip_shape}}. Also, the fact that Eq.~(1) accurately describes the potential around circular holes, while many of the holes considered in our experiments deviate significantly from a circular shape, can be the origin of a poor fit with Eq.~(1). {We note that the deviation of the potential from Eq.~(1) due to noncircularity is most prominent close to the hole. Farther away from the hole, the potential is expected to approach again the profile of a circular hole. }}

{For a better description of the data close to the hole edges, we therefore also performed fits with a second, more general (`relaxed') fitting function, namely
\begin{align}
    V_\mathrm{dipole}=A/(x-x_0)+B\cdot (x-x_0)+C,
    \label{eq:relaxed_fit}
\end{align}
where $x_0$, $A$, $B$ and $C$ are fit parameters without further restrictions. We fitted this equation to each side of the hole separately {and independently}. Importantly, when considering a circular hole, both fits [\textit{i.e.}, Eqs.~(\ref{eq:relaxed_fit}) and (1)] give the same result. For noncircular holes, however, the strict fit with Eq.~(1) can over- or underestimate the {potential at the hole edge,}
depending on the details of the hole shape. 
For such holes, the relaxed fit typically describes the data better in a purely mathematical sense, \textit{i.e.} with less residual error.}

{We stress that both fit routines have their advantages and disadvantages, such that it is \textit{a priori} not clear which allows the overall more reliable determination of the resistivity dipole moments of a given set of holes---and therefore, {for} an unbiased analysis {we use} both fit routines.
In general, the advantage of the relaxed fit, \textit{i.e.}, a smaller residual error near the hole edge,
are offset by its greater sensitivity to noise and measurement artifacts, including thermovoltages and other systematic errors such as the spatial broadening of the potential close to the hole edge that is shown in Fig.~\ref{fig:Sup3}c and discussed in section~\ref{sec:tip_shape}. 
Also, the relaxed fit is more sensitive to the potential from higher multipole moments that may be present close to the hole edge because of deviations from the circularity---when trying to extract the resistivity dipole, such higher multipoles are a distraction. 
Conversely, advantages of the strict fit are: 
A solid grounding in the physics of dipoles, a faithful adherence to the $1/r$ far field even for noncircular holes, and resulting from that, less sensitivity to multipole moments, and potential broadening and other artifacts close to hole edges. 
Furthermore, and unlike the relaxed fit, the strict fit enforces continuity of the potential through the hole, since both branches $x\gtrless0$ are fitted simultaneously. 
}

{
In the following, we applied both fitting routines to data set B, with a direct comparison between the two fit approaches for one hole from data set B is displayed in Fig.~\ref{fig:Sup5}.
We note that for the relaxed fit generally $V_\mathrm{dipole}(-a)\neq - V_\mathrm{dipole}(+a)$. We nevertheless \textit{define} a symmetric dipole potential amplitude $2|V_\mathrm{dipole}(a)|\equiv V_\mathrm{dipole}(-a)- V_\mathrm{dipole}(+a)$, \textit{i.e.}, in our terminology $|V_\mathrm{dipole}(a)|$ is not the modulus of $V_\mathrm{dipole}(+a)$. A similar argument applies to $\tilde V_\mathrm{D}(x)$ from the resistor network calculation, which in general is also not symmetric with respect to $x=0$.
When applied to data set A, however, the relaxed Eq.~(\ref{eq:relaxed_fit}) did not result in proper fits, because its flexibility tended to respond too sensitively to the significant deviations of the measured potential from the dipole shape that are caused by the presence of thermovoltages in {the} potential data (see section~\ref{sec:thermovoltage}). 
We therefore did not include the results of fits of data set A with Eq.~(\ref{eq:relaxed_fit}) in {the following}. 
}

{From each fit, we extracted the experimental dipole potential amplitude $2|V_\mathrm{dipole}(a)|$ {and compare it} to the analytical diffusive dipole potential amplitude  $2|V_\mathrm{D}(a)|$ from Eq.~(\ref{eq2:dipoleMoment_diffusive}), evaluated at the threshold-determined hole size. 
For the large{r holes (Fig.~\ref{fig:Sup2}a,c)} we see {a good overall} agreement between $V_\mathrm{D}(x)$ (dotted teal {curve}) and $V_\mathrm{dipole}(x)$ (solid black {curve}).
For the small{er holes (Fig.~\ref{fig:Sup2}b,d)}, however, $2|V_\mathrm{D}(a)|$ 
is significantly smaller than $2|V_\mathrm{dipole}(a)|$. 
While this is a first indication that the resistivity dipole moment at this hole cannot be described by diffusive transport theory, we note that the analytical description $V_\mathrm{D}$ considers a circular hole. 
Thus, it is \textit{a priori} not clear if the discrepancy between $V_\mathrm{D}$ and $V_\mathrm{dipole}$ is just a consequence of the hole shape.

The potential maps {resulting from the resistor network calculations} $\tilde V_\mathbf{D}(x,y)$ (bottom right insets in Fig.~\ref{fig:Sup2}) can be directly compared with the experimental potentiometry data (bottom left insets). 
For the cross section through the large hole, the symmetrized $2|\tilde V_\mathrm{D}(a)|\equiv \tilde V_\mathrm{D}(-a)-\tilde V_\mathrm{D}(+a)$ is in {good} agreement with both $2|V_\mathrm{D}(a)|$ and $2|V_\mathrm{dipole}(a)|$ (Fig.~\ref{fig:Sup2}c), confirming that the transport around this defect can be described by the diffusive transport theory. In contrast, for the small hole, {we find $2|\tilde V_\mathrm{D}(a)|\approx 2|V_\mathrm{D}(a)|\approx 0.8 \times 2|V_\mathrm{dipole}(a)|$, revealing that the smaller hole size rather than its shape is responsible for the difference between the expectation $2|V_\mathrm{D}(a)|$ for a diffusive dipole and the experimental result $2|V_\mathrm{dipole}(a)|$.}

{
\subsection{Transition from diffusive to Landauer regime}
}
To check whether this {size-induced} deviation is systematic, we evaluated 35 different holes of widely different sizes (examples are plotted in the Appendix A).
{By n}ormalizing $|V_\mathrm{dipole}(a)|$ by the current density {in 2D}, $j=\sigma E_0$, we further took into account variations in the current density between different measurements, and thereby determined the resistivity dipole $\rho_\mathrm{dipole} = |V_\mathrm{dipole}(a)|/j$: 
{To extract the resistivity dipoles from the measured dipole potentials, we remember that in the diffusive limit the total field within the defect is $2E_0$ (see above). After subtraction of the linear background potential of the driving field ($V=-E_0x$) the remaining field within the defect is thus $E_0$, with the result that the initial driving field can be expressed as $E_0=|V_\mathrm{dipole}(a)|/a_\parallel$. Then, from $p_\mathrm{D}(a)=a_\parallel a_\perp E_0=\rho_\mathrm{dipole}^\mathrm{D} a_\perp j$ follows $\rho_\mathrm{dipole}^\mathrm{D}=|V_\mathrm{dipole}(a)|/j$. Similarly, from $p_\mathrm{L}=a_{0,\parallel} a_\perp E_0=\rho_\mathrm{dipole}^\mathrm{L} a_\perp j$ we obtain $|V_\mathrm{dipole}(a)|/j=(a_\parallel/a_{0,\parallel})\rho_\mathrm{dipole}^\mathrm{L}$. Since in the Landauer limit $a_\parallel$ is not relevant, it can be replaced by the corresponding threshold size $a_{0,\parallel}$. Thus, plotting the experimental $|V_\mathrm{dipole}(a)|/j=\rho_\mathrm{dipole}$ vs. $a$, we can easily distinguish between the diffusive and Landauer behavior.}
{We note that} $\rho_\mathrm{dipole} =p/(a_\perp j)$ is the resistivity dipole moment normalized to the transverse hole size (\textit{i.e.}, scattering cross section) and the current density. At the same time, using Eqs.~(\ref{eq1:potentialDipole}) and (\ref{eq2:dipoleMoment_diffusive}), we converted $|\tilde V_\mathrm{D}(a)|$ of the resistor network calculation to an effective {(lateral)} hole size $a^*\equiv |\tilde V_\mathrm{D}(a)|/E_0$, which reflects {the size of} a circular hole in the diffusive limit that would produce the given $|V_\mathrm{dipole}(a)|$. 

\begin{figure}[bt]
\begin{center}
\includegraphics[width=\columnwidth]{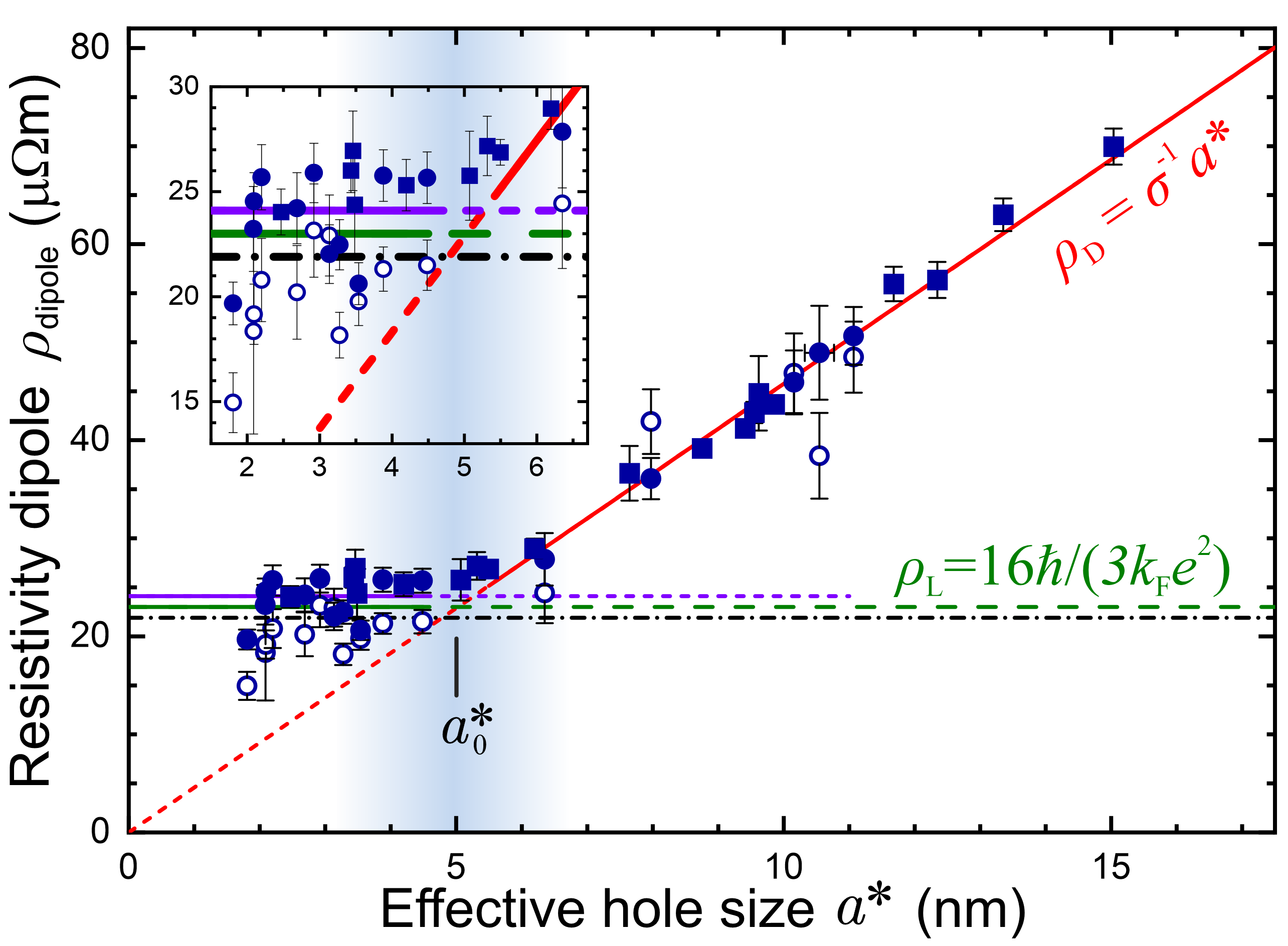}
\end{center}
\caption{\label{fig4} \textbf{Transition from diffusive to Landauer resistivity dipoles.} 
Experimental resistivity dipole $\rho_{\rm dipole} = |V_{\rm dipole}(a)|/j$ as function of the effective hole size $a^* \equiv |\tilde V_\mathrm{D}(a)|/E_0$  for 
{data set A (squares) and data set B (circles).
Solid symbols are the result of point-symmetric fits {to} the dipole potential with Eq.~(\ref{eq1:potentialDipole}) (strict fit routine), while empty symbols for data set B result from separate fits to both sides of the dipole potential [{Eq.~\eqref{eq:relaxed_fit}}, relaxed fit routine].}
The solid red line corresponds to the diffusive limit. {The solid green and purple lines are fits to all data below $a^*_0$ and only filled symbols below $a^*_0$, respectively. 
The black {dash-dotted} line is the approximate minimum value of the Landauer dipole based on Refs.~\onlinecite{bian2009electronic, bian2014first}.}
{Inset: Zoom into the transition region.}
}
\end{figure}

In Fig.~\ref{fig4}, we plot the experimental $\rho_\mathrm{dipole}$ vs.~$a^*${, including resistivity dipoles extracted from data sets A and B and using the two different fit routines.} For $a^*>a_0^*$, where $a_0^* \approx 5$\,nm, the data points fall on a linear slope, following $\rho^\mathrm{D}_\mathrm{dipole}(a^*)=\sigma^{-1}a^*$, proving that scattering at defects of this size can be described by diffusive transport theory. In contrast, for $a^*<a_0^*$ {all} resistivity dipoles are above the diffusive slope{, with the sample average more than $\sim 6.7$ standard errors away from the null hypothesis that diffusive transport theory also applies for $a^*<a_0^*$. The confidence level with which this hypothesis can be rejected is thus $>99.995$\% (see Appendix B). The data points instead tend towards a constant. This is {the smoking gun} signature of Landauer resistivity dipoles. 

{Fig.~\ref{fig4} {further} reveals that data points from the relaxed fit routine show larger error bars and a more significant scatter{, despite being more flexible in fitting the data close to the hole edges.} 
{The smaller scattering of data points for the strict fitting routine suggests that the asymptotic potential far away from a hole allows a more reliable determination of the latter's resistivity dipole moment than the potential close to the hole edge, where it can be affected by higher multipole moments and measurement artifacts, as discussed above in section \ref{sec:fitting}.}
Thus, the fit curves of the strict routine describe the real potential (and therefore the resistivity dipole moment) better than the data points in this regime, which are subject {to} systematic errors as discussed above. It is, however, important to note that both fit routines support the main conclusion of our work: For holes below a critical size, both routines very clearly show the deviation from the diffusive limit and the trend towards the constant Landauer dipole, with only a small difference between the constant value they tend towards.
}

{Photoemission measurements of {similar} Bi films in Refs.~\onlinecite{bian2009electronic, bian2014first} indicate that $k_{\rm F}\lesssim 1\,\mathrm{nm}^{-1}$, which according to Eq.~(\ref{eq3:dipoleMoment_Landauer}) results in a lower bound for the Landauer dipole $\rho^\mathrm{L}_\mathrm{dipole}\gtrsim22\,$\textmu$\mathrm{\Omega m}$ (black {dash-dotted} line in Fig.~\ref{fig4}). 
{However, this value should be taken with caution, because the underlying photoemission measurements carry significant uncertainty.
Nonetheless}, the value which our data points approach  for  $a^*<a_0^*$ is very close to this limit. 
{Although this transition is expected to be only complete once $a^*\ll a^*_0$, the data indicate that it proceeds quickly once $a^*$ falls below the threshold $a^*_0$.}
To estimate the value of the Landauer dipole, we fit all} data points for $a^*<a^*_0$ with a horizontal line, obtaining $\rho^\mathrm{L}_\mathrm{dipole}=(23 \pm 3)\,$\textmu$\mathrm{\Omega m}$ (green dashed line in Fig.~\ref{fig4}).
From 
{
Eq.~\ref{eq:rho_L}}, we obtain the Fermi wave vector in the Bi film as { $k_\mathrm{F}=(0.95 \pm 0.07)\,\mathrm{nm}^{-1}$}. This value is in agreement with the {$k_{\rm F}\lesssim1\,\mathrm{nm}^{-1}$} observed in Refs. \onlinecite{bian2009electronic, bian2014first}, although theoretical calculations predict a larger $k_{\rm F} \approx 3$~nm$^{-1}$ (Ref. \onlinecite{Yaginuma2007, bian2014first, lu2015topological}). 
This discrepancy between theory and experiment is most likely caused by a combination of charge doping and disorder, both of which may originate from substrate interactions. 

Finally, from the hole size { $a^*_0\approx 5\,\mathrm{nm}$} at which the transition between the diffusive and ballistic regimes occurs {in our data}, { \textit{i.e.}, where the two lines in Fig.~\ref{fig4} intersect,} we estimate a mean free path of
{
\begin{align}
    \lambda=3\pi a^*_0/8 = (6 \pm 1)\,\mathrm{nm}. \label{eq:lambda}
\end{align}
}This value is in the range of $\lambda \approx 3\,\mathrm{nm}$~to~$10\,\mathrm{nm}$ reported in the literature for Bi films on Si(111)-$7\times 7$ \cite{Feenstra1998, Hirahara2007, Briner1996}. 
We further note that the theoretical conductivity of the 2DEG in the Drude model is {
\cite{Datta1997}
\begin{align}
    {\sigma=(e^2/h)k_\mathrm{F}\lambda=(0.22\pm0.04)\rm\,mS/\square}, \label{eq:sigma}
\end{align}
}where we used the values for $k_\mathrm{F}$ and $\lambda$ extracted from our analysis. 
This result is in excellent agreement with our {independently measured} sheet conductivity $\sigma=(0.22\pm0.01)\rm\,mS/\square$, confirming the consistency of our analysis.

{In comparison to recent studies of resistivity dipoles around nanoscale pits in graphene on SiC \cite{markovic2025intermediate}, our work reveals the diffusive-to-ballistic crossover by systematically varying the defect size and explicitly accounting for the influence of defect geometry.
Nonetheless, the results on graphene/SiC are consistent with our work, considering the graphene's larger mean free path ($\lambda\approx35$\,nm at 90\,K) and smaller $k_\mathrm{F} = 0.72\rm\,nm^{-1}$:
The corresponding transition length (for circular holes) is $a_0^*\approx39\rm\,nm$ [Eq.~(\ref{eq:lambda})], \textit{i.e.}, much larger than in our Bi films. Yet, the resulting Landauer dipole in graphene $\rho_\mathrm{dipole}^\mathrm{L}=16\hbar/(3k_\mathrm{F}e^2)\approx30$\,\textmu$\Omega$m [Eq.~(\ref{eq:rho_L})] is only slightly higher than in our Bi films, due to the larger graphene conductivity:
The theoretical sheet conductivity of graphene includes an additional factor of 2, due to valley degeneracy, resulting in $\sigma=2(e^2/h)k_\mathrm{F}\lambda\approx2\rm\,mS/\square$ [Eq.~(\ref{eq:sigma})], which is consistent with the measured conductivity in that work \cite{markovic2025intermediate}.
}

{
\section{Conclusion}}
In conclusion, we have provided compelling evidence for the transition between diffusive and ballistic resistivity dipoles at defects in ultra-thin Bi films. The moderate mean free paths in our samples proved amenable to identifying defects in both limits, allowing to unambiguously pinpoint the transition and consistently extract crucial material parameters of our Bi films: 
the mean free path $\lambda$ from the defect size at which the transition {to the ballistic scattering regime} occurs and the Fermi wave vector $k_\mathrm{F}$ from the saturation value of the {Landauer residual resistivity dipole} at small defect size. Notably, our resistor network calculations take into account the significant influence of the irregular defect shapes in the diffusive regime, allowing us to exclude the latter as a source of the observed deviations from the analytical solution for circular holes. 
{In the future, this approach could be extended with lattice-Boltzmann theory \cite{markovic2025intermediate}, to accurately model the mixed diffusive-ballistic dipoles around arbitrarily shaped defect in the transition region.}
We foresee that the particularly clear picture of the resistivity dipoles reported here will facilitate further investigations in at least { three} directions:  First, because a decrease in temperature will shift the transition between the diffusive and ballistic regimes to larger defects, a detailed study of temperature-dependent transport (including means free paths) will become possible. Such low-temperature STP experiments also promise to resolve the transition in even more detail, because of further improvements in experimental resolution \cite{Luepke2015} and access to more detailed information of the Fermi surface \cite{Wexler1966, Allen2006, Popov1963}. 
{
Second, our experimental data may be used for benchmarking the development of theoretical models that describe the transition between the transport regimes.
Third}, our clear identification scheme of the ballistic regime will greatly benefit the search for current-induced oscillations in the carrier density \cite{zwerger1991exact, kramer2012quantum}, which are expected around defects in addition to the Landauer resistivity dipole and the well-known (and often observed) Friedel oscillations at zero current \cite{friedel1954electronic}.\\

\section{Acknowledgements}
This work was supported by the German excellence cluster ML4Q (Matter and Light for Quantum Computing).
S.K. acknowledges the Polish National Agency for Academic Exchange NAWA (Bekker programme III edition on 23.05.2021).
F.L.\ and F.S.T.\ acknowledge the financial support by the Bavarian Ministry of Economic Affairs, Regional Development and Energy within Bavaria’s High-Tech Agenda Project ``Bausteine f\"ur das Quantencomputing auf Basis topologischer Materialien mit experimentellen und theoretischen Ans\"atzen'' (Grant No.\ 07 02/686 58/1/21 1/22 2/23). 
F.L.\ acknowledges funding by the Deutsche Forschungsgemeinschaft (DFG, German Research Foundation) within the Research Unit FOR 5242 (Project No.\ 449119662), as well as the Emmy Noether Programme (Project No.\ 511561801). 
F.S.T.\ acknowledges funding by the Deutsche Forschungsgemeinschaft (DFG, German Research Foundation) through Coordinated Research Center CRC 1083, project ID 223848855.

{
B.V., F.S.T, and F.L. {conceived and designed} the research.
B.V., I.M., F.S.T., and F.L. supervised the project.
S.K., D.K., V.C., and F.L. conducted the experiments. 
S.K., D.K., J.H. T.B., V.C., B.V. and F.L. analyzed the data.
S.K., V.C., B.V., F.S.T., and F.L. wrote the manuscript, and all authors commented on it.
}

{\section{Data availability}
The data that support the findings of this article, as well as the analysis scripts used to produced the shown figures, will be made openly available at \textit{J\"ulich Data} upon publication.
}

%

\setcounter{figure}{0} 
\renewcommand{\thefigure}{A\arabic{figure}}

{
\section*{Appendix}
\subsection{Potential data for additional dipoles}\label{Appendix:A}
Fig.~\ref{fig:Sup6} shows the same analysis as presented in Fig.~\ref{fig:Sup2} for additional holes belonging to data set B.
\begin{figure*}[bt]
\begin{center}
	\includegraphics[width=2\columnwidth]{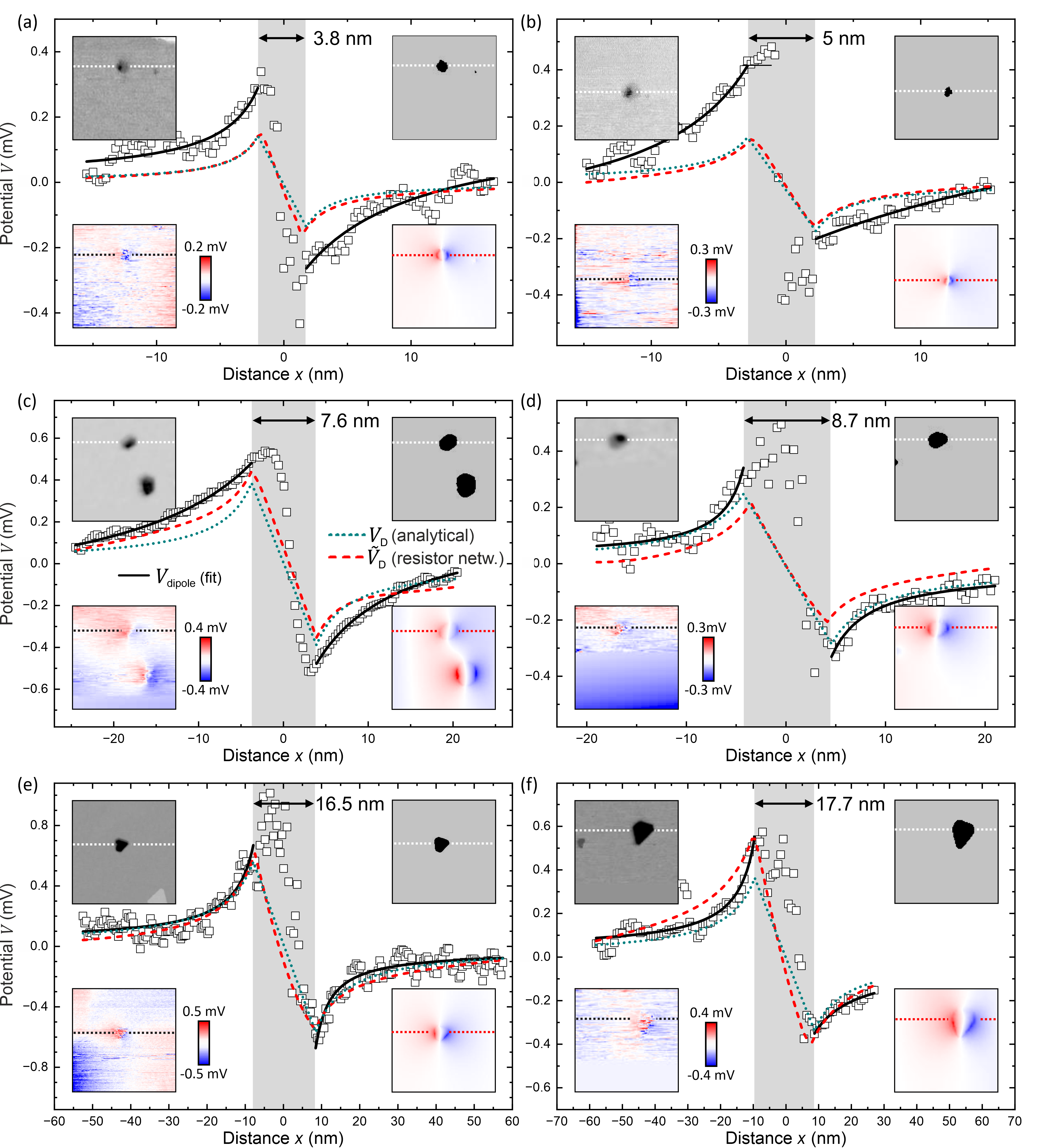}
 \end{center}
   \caption{\label{fig:Sup6} 
  \textbf{Data for additional holes from data set B.} {The fits (black solid lines) were carried out with Eq.~\ref{eq:relaxed_fit} separately for $x<0$ and $x>0$}. For the small holes with $a \lesssim 5\,$nm in panels (a) to (d), the dipole potential amplitudes of the corresponding resistor network calculations (defined by extrema of red dashed lines) are smaller than the experimentally measured ones (defined by extrema of black squares and black solid lines). For the larger holes with $a \gtrsim 5\,$nm in panels (e) and (f), the resistor network calculations coincide with to the experimental data, showing that the larger holes are in the diffusive transport regime.
  { The current densities in panels (a) to (f) are $j=(9.55\pm 0.93)\rm\,A/m$, $(8.8\pm 2.4)\rm\,A/m$, $(12.4 \pm 2.7)\rm\,A/m$, $(8.04\pm 0.97)\rm\,A/m$, $(7.90 \pm 0.56)\rm\,A/m$, $(11.22 \pm 0.86)\rm\,A/m$.}
    }
\end{figure*}
}

{
\subsection{Hypothesis test}\label{Appendix:B}
To estimate the statistical significance of the findings displayed in Fig.~\ref{fig4}, we employed a one-sample student's $t$ test. We tested the null hypothesis $H_0$ that the experimental data points from the relaxed dipole fits for small holes (open circles in Fig.~\ref{fig4}) follow the diffusive behavior (red line in Fig.~\ref{fig4}) against the alternative hypothesis $H_1$ that they deviate to one side from diffusive transport. 
For this purpose, we calculated the residuals $r_i=\rho_i-\tilde\rho_i$, where $\rho_i$ are the data points for $a^*<a^*_0$ and $\tilde\rho_i$ are the theoretical diffusive resistivity dipoles expected for the corresponding $a^*$ values. If, due to the central limit theorem, in the parent population all $\rho_i$ are normally distributed with standard deviation $\sigma$ (which in principle could be determined from an infinite sample that approaches the parent population), then also the residuals $r_i$ will be, since the $\tilde{\rho}_i$ only shift the expectation value, but do not change the distribution. Evidently, under the null hypothesis the residuals $r_i$ have a common expectation value $\mu_0=0$.  We then calculated the average of the residuals
\begin{equation}
    \bar{r}=\frac{1}{n}{\sum_{i=0}^{n} r_i},
    \label{eq:average}
\end{equation}
where $n=11$ is the number of data points in our case.
Under the null hypothesis, also the average $\bar r$ is normally distributed around zero,  but with standard deviation $\sigma/\sqrt{n}$. 
Since $\sigma$ (which belongs to the parent population) is not known, 
it is common to use in its place the \textit{empirical} standard deviation 
\begin{equation}
    s=\sqrt{\frac{1}{n-1}\sum_{i=0}^{n} (r_i-\bar{r})^2}
    \label{eq:empirical_sd}
\end{equation}
to estimate the standard deviation of the finite sample from the sample itself. We then used the student's $t$, given by 
\begin{equation}
    T=\frac{\bar r -\mu_0}{s/\sqrt{n}},
    \label{eq:T}
\end{equation}
as test statistics. Here, $\sigma_{\rm SE}=s/\sqrt{n}$ is the standard error of the sample. $T$, instead of being normally distributed, follows the $t$-distribution with $\rm{df}=n-1$ degrees of freedom. The $t$-distribution  is centered at zero.  For large $n$, it approaches the normal distribution, for small $n$ is has heavy tails compared to the Gau{\ss} distribution. The actual value of $T$ for our sample of 11 data points is $t=6.7286$ ($\bar r= 6.67\,$\textmu$\mathrm{\Omega m}$,  $s=3.29\,$\textmu$\mathrm{\Omega m}$), which means that the measured sample average $\bar r$ is $\sim6.7$ standard errors $\sigma_{\rm SE}$ away from the average $\mu_0$ under the null hypothesis ($\bar r\approx 6.7 \sigma_{\rm SE}$). This result can be discussed in terms of either the significance level $\alpha$ or the so-called $p$ value. The significance level is the upper bound for the probability that the null hypothesis is erroneously rejected, and defined as 
\begin{equation}
    \alpha=P(T\geq t_{\rm c}| H_0)\quad \quad\quad \text{for}\,\, t_{\rm c}>0, 
\label{eq:significance_level}
\end{equation}
\textit{i.e.}, as the conditional probability $P$ that, if $H_0$ was true, we would find $T\geq t_{\rm c}$, where $T$ is the test statistic [Eq.~(\ref{eq:T})]  and $t_{\rm c}$ the critical value of $T$ beyond which (\textit{i.e.}, for $T>t_{\rm c}$) we would reject $H_0$. For a given distribution, $t_{\rm c}$ follows from $\alpha$ and the number of degrees of freedom ${\rm df}=n-1$. In our case of  a one-sided test using the student's t-distribution, $\alpha=0.00005$ and ${\rm df}=10$ yield $t_{\rm c}(\alpha=0.00005,{\rm df}=10)=6.211$. Because of the chosen significance level, the probability of erroneously rejecting $H_0$ although it is true is 0.005\% at most. Since the measured $t$ of our sample is larger than the critical one ($t>t_{\rm c}$), we may therefore not only reject the null hypothesis, but also safely do so. We can also discuss our result in terms of the $p$ value, which is the smallest possible significance level at which the null hypothesis would still be rejected. It is defined as 
\begin{equation}
    p=P(T\geq t| H_0)\quad \quad\quad \text{for}\,\, t>0
    \label{eq:p_value}
\end{equation}
where $t$ is the measured $T$ value of a sample. Note that from Eqs.~(\ref{eq:significance_level}) and (\ref{eq:p_value}) follows
\begin{equation}
    p<\alpha \Leftrightarrow t>t_{\rm c}\quad \quad\quad \text{for}\,\, t,\,t_{\rm c}>0.
    \label{eq:alpha_p}
\end{equation}
In the present case with $t=6.729$ (rounded) and ${\rm df}=10$, we obtain $p=0.00002586$. This $p$ value means that if $H_0$ was true, only 0.002586\% of randomly drawn samples with $n=11$ (\textit{i.e.}, only one in approximately 38,000) would reject it. Since our sample rejects the null hypothesis with very large statistical significance (given by the very small $\alpha=0.00005$, see above), and only very few samples would do this if $H_0$ was true (as the small $p$ value tells us), we can again safely conclude that the null hypothesis is false.
} 

{We also applied the one-sided student's $t$ test to check the alternative null hypothesis $H_0'$ that \textit{all} 27 data points for hole of size $a^*<a^*_0$ lie on the green line in Fig.~\ref{fig4}. From Eqs.~\ref{eq:average} and \ref{eq:empirical_sd} we obtained $\bar r'=-0.4299\,$\textmu$\mathrm{\Omega m}$ and  $s'=3.018\,$\textmu$\mathrm{\Omega m}$. This yields $t'=-0.74017$, \textit{i.e.}, 
$|\bar r'|\approx 0.74\sigma_\mathrm{SE}$. The corresponding $p$ value for a one-sided student's $t$-test with ${\rm df}=26$ is $p'=0.233$. Thus, for a measured $t$ value slightly above $|t'_{\rm c}|=0.74017$, we would reject $H_0'$, but assuming that $H_0'$ was true, $\sim 23$\% of all samples would do so erroneously. Our rejection would therefore not be very safe.  Conversely, for a measured $t$ value slightly below $|t'_{\rm c}|=0.74017$, we would accept $H_0'$, but again this would not be a strong statement, because if $H_0'$ was true, only $\sim 77$\% of the samples would accept it correctly. We thus conclude that the null hypothesis $H_0'$ according to which all data points lie on the green line in Fig.~\ref{fig4} is consistent with our data, but the truth of this  hypothesis cannot be asserted with high statistical significance on the basis of our data.  This is actually not surprising, because of the significant spread of data values from the two different fit routines.
{Furthermore, }because the convergence to the Landauer resistivity dipole is not completed for holes that are not much smaller than the transition length $a^*_0$ and, there may still be systematic deviations from the green curve in the range of our data. 
In summary, our data (i) support the occurrence of a cross-over away from the diffusive resistivity dipoles at $a^*_0$  with very high statistical significance, (ii) are fully consistent with a cross-over into a regime with size-independent Landauer resistivity dipoles, and (iii) the amplitude of the dipoles below $a^*_0$ is consistent with the minimum Landauer dipole amplitude based on Ref.~\onlinecite{bian2014first}.} 

\end{document}